\documentclass{mem}
\usepackage{natbib}\usepackage{txfonts}\usepackage{balance}
\usepackage{graphicx}
\usepackage[a4paper]{hyperref}
\idline{}{1}
\begin{document}
\title{Probing the Local Bubble\\ with\\ Diffuse Interstellar Bands (DIBs)}
\author{Jacco Th.\ van Loon\inst{1}
\and A.\ Farhang\inst{2}
\and A.\ Javadi\inst{2}
\and M.\ Bailey\inst{3}
\and H.G.\ Khosroshahi\inst{2}}
\institute{Lennard-Jones Laboratories, Keele University, ST5 5BG, UK\\
\email{j.t.van.loon@keele.ac.uk}
\and School of Astronomy, IPM, Tehran, Iran
\and Astrophysics Research Institute, Liverpool John Moores University, UK}
\authorrunning{van Loon et al.}
\titlerunning{Local Bubble with DIBs}
\abstract{The Sun lies in the middle of an enormous cavity of a million degree
gas, known as the Local Bubble. The Local Bubble is surrounded by a wall of
denser neutral and ionized gas. The Local Bubble extends around 100 pc in the
plane of Galaxy and hundreds of parsecs vertically, but absorption-line
surveys of neutral sodium and singly-ionized calcium have revealed a highly
irregular structure and the presence of neutral clouds within an otherwise
tenuous and hot gas. We have undertaken an all-sky, European-Iranian survey of
the Local Bubble in the absorption of a number of diffuse interstellar bands
(DIBs) to offer a novel view of our neighbourhood. Our dedicated campaigns
with ESO's New Technology Telescope and the ING's Isaac Newton Telescope
comprise high signal-to-noise, medium-resolution spectra, concentrating on the
5780 and 5797 \AA\ bands which trace ionized/irradiated and neutral/shielded
environments, respectively; their carriers are unknown but likely to be large
carbonaceous molecules. With about 660 sightlines towards early-type stars
distributed over distances up to about 200 pc, our data allow us to
reconstruct the first ever 3D DIB map of the Local Bubble, which we present
here. While we confirm our expectations that the 5780 \AA\ DIB is relatively
strong compared to the 5797 \AA\ DIB in hot/irradiated regions such as which
prevail within the Local Bubble and its walls, and the opposite is true for
cooler/shielded regions beyond the confines of the Local Bubble, we
unexpectedly also detect DIB cloudlets inside of the Local Bubble. These
results reveal new insight into the structure of the Local Bubble, as well as
helping constrain our understanding of the carriers of the DIBs.
\keywords{
ISM: atoms --
ISM: bubbles --
ISM: individual objects: Local Bubble --
ISM: molecules --
ISM: structure --
local interstellar matter}}
\maketitle{}
\section{Introduction}

\begin{table*}
\caption{The UK--Iran ultra-deep all-sky survey of DIBs in and around the
Local Bubble.}
\label{survey}
\begin{center}
\begin{tabular}{|l|c|c|c|}
\hline
                      & Southern     & Northern     & Combined   \\
\hline
telescope             & 3.6m NTT     & 2.5m INT     & $\sim3$m   \\
observatory           & ESO La Silla & ING La Palma & $\pm29^\circ$ latitude \\
period                & 2011--2012   & 2011--2013   & 2011--2013 \\
$\lambda/\Delta\lambda$ & 5500       & 2000         & 2000--5500 \\
signal-to-noise       & 1000--2000   & 1000--2000   & 1000--2000 \\
\# sight-lines        & 238          & 432          & 655 unique \\
types                 & O,B          & B,A          & O,B,A      \\
reference             & \citet{Bailey2015b}
                      & \cite{Farhang2015a,Farhang2015b}
                      & Farhang et al.\ (2016) \\
\hline
\end{tabular}
\end{center}
\end{table*}

The Solar System currently finds itself in an area within the Milky Way Disc
of relatively low interstellar density called the ``Local Bubble''
\citep{Fitzgerald1968}. It is not a void, though, and while filled largely
with million-degree plasma \citep{Snowden2015} it is far from homogeneous and
contains denser, cooler ``local fluff'' \citep{Redfield2000}. The structure of
the Local Bubble has been mapped in three dimensions using stars with known
distances from parallax measurements, through Na\,{\sc i} and Ca\,{\sc ii}
spectral line absorption by neutral and ionized gas
\citep{Lallement2003,Welsh2010} and photometric reddening by dust
\citep{Lallement2014}. Given the importance of the local interstellar
environment for the effectiveness of the heliosphere to protect life on Earth,
and to better understand the dynamics of the multi-phase interstellar medium
(ISM) in spiral galaxies, novel ways of mapping the Local Bubble remain highly
desirable.

A tracer of weakly-ionized gas, the diffuse interstellar bands (DIBs) that are
ubiquitous in the optical spectra of stars have been known for almost a
century \citep{Heger1922}. They show a huge variety in widths, strengths and
shapes, yet their carriers remain elusive \citep{Herbig1995,Sarre2006}. It is
likely that the carriers are carbonaceous molecules, possibly charged, and
they could include members of the family of fullerenes
\citep{Foing1994,GarciaHernandez2013,Campbell2015}.

DIBs are increasingly recognised as powerful tools to map the ISM, since the
pioneering work by \citet{vanLoon2009} who used the $\lambda$5780 \AA\ and
$\lambda$5797 \AA\ DIBs to map the extra-planar gas in front of hundreds of
metal-poor, blue horizontal branch stars in the largest Galactic globular
cluster, $\omega$\,Centauri. These maps revealed parsec-scale structure, and
anti-correlated behaviour of these two DIBs confirming that the $\lambda$5780
\AA\ and $\lambda$5797 \AA\ DIBs trace different interstellar conditions. In
a subsequent study, \citet{vanLoon2013} mapped the strong $\lambda4428$ \AA\
and $\lambda$6614 \AA\ DIBs using the spectra of about 800 stars within the
Tarantula Nebula in the Large Magellanic Cloud (LMC). These maps showed the
presence of DIBs in both the diffuse ionized and neutral ISM but their
disappearance near strong sources of UV radiation; this behaviour resembles
that of the equally unidentified infrared emission features (commonly
attributed to polycyclic aromatic hydrocarbons). Due to the Doppler shift of
the LMC, the foreground component arising in the local Galactic ISM could be
mapped separately, again showing pc-scale structure in the DIB carriers,
excitation conditions, or both, but this time at even higher Galactic
latitude.

\citet{Kos2014}, \citet{Lan2015} and \citet{Zasowski2015} utilized the SDSS,
RAVE and APOGEE surveys to map the kiloparsec-scale distribution of DIBs
across the Galactic Disc and Halo. To map the low column density material in
and around the Local Bubble requires much higher signal-to-noise than what is
typical in stellar and extra-galactic spectroscopic surveys, and hence we
initiated an ultra-deep, all-sky survey dedicated to the detection of DIBs at
the sub-percent level.

\section{The UK--Iranian ultra-deep all-sky Local Bubble DIBs survey}

\begin{figure*}[t!]
\resizebox{\hsize}{!}{\includegraphics[clip=true]{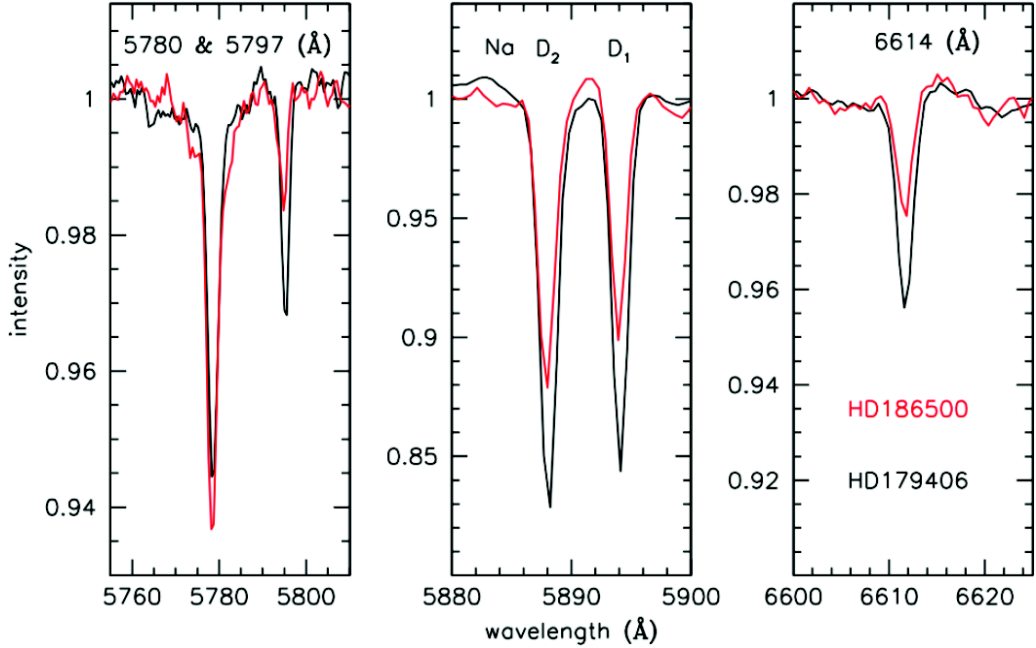}}
\caption{\footnotesize
Difference in behaviour between the $\lambda$5780 \AA\ DIB ({\it leftmost})
and the $\lambda$5797 and 6614 \AA\ DIBs and the Na\,{\sc i}\,D lines; the
higher-latitude sight-line towards HD\,186500 is an example of the $\sigma$
conditions in which the $\lambda$5780 \AA\ DIB is relatively strong, whereas
the opposite, $\zeta$ conditions are found towards HD\,179406. Expanded on
\cite{Bailey2015b}.}
\label{sigmazeta}
\end{figure*}

Starting with a survey in the Southern hemisphere (including targets up to
about $+10^\circ$ declination), with the New Technology Telescope (NTT) at the
European Southern Observatory (ESO) at La Silla, Chile \citep{Bailey2015b}, it
was extended to an all-sky survey by a complementary Northern hemisphere
component conducted with the Isaac Newton Telescope (INT) at La Palma, Canary
Islands \citep{Farhang2015a,Farhang2015b} -- see table~\ref{survey}. By virtue
of the prevailing climatic conditions near the tropical circles the La Silla
and La Palma observatories are at almost exact opposite sides from the Earth's
equator!

Targets were chosen to be predominantly early-type stars (O and B type), but
additional, later-type (A type) stars were included in the Northern survey to
improve coverage -- most targets are naked-eye stars in order to reach the
very high signal-to-noise ($>1000$). The spectra of the cooler stars were
fitted with synthetic model atmosphere spectra to separate interstellar from
photospheric absorption, and even in the spectra of the hotter stars some
interfering features had to be fitted simultaneously and removed from the
interstellar band. All stars have known parallaxes (and proper motions), and
are mostly within a few hundred pc from the Sun; their three-dimensional
spacings are typically 10--40 pc. In total, 655 unique sight-lines were
probed, of which 15 were observed in both hemisphere survey components
(reassuringly, their measurements were found to agree very well).

An example of two sight-lines probing different environments is presented in
figure~\ref{sigmazeta}. The $\lambda$5797 \AA\ DIB, and also the $\lambda$6614
\AA\ DIB, generally traces the neutral gas that also dominates the column
density probed by the Na\,{\sc i}\,D lines (and reddening), but the
$\lambda$5780 \AA\ DIB exhibits large variations, likely depending on the
level of irradiation and/or the contact with hot gas. This dichotomy is well
known since \citet{Krelowski1988} compared the spectrum of $\zeta$\,Ophiuchi
(relatively strong $\lambda$5797 \AA\ DIB, henceforth called $\zeta$ clouds)
with that of $\sigma$\,Scorpii (relatively strong $\lambda$5780 \AA\ DIB,
henceforth called $\sigma$ clouds) -- despite their near-identical reddening.

Indeed, we found a large $\lambda$5780/$\lambda$5797 DIB ratio within the
Local Bubble and generally in the extra-planar gas. The ratio is often low for
sodium equivalent widths of $EW($Na\,{\sc i}\,D$_2)>0.2$ \AA, which
corresponds to column densities where the sodium D lines become affected
significantly by saturation. For denser columns the sodium equivalent width
increases more easily as a result of the increasing multiplicity of clouds
along the sight-line; higher resolution studies confirm that is the case.

Likewise, the correlations between DIBs seem not as simple at low column
densities than at moderate-to-high column densities. We thus conjecture that
the good -- at times near perfect \citep{McCall2010} -- correlations seen
between some DIBs and between DIBs and other tracers such as atomic lines or
reddening \citep{Friedman2011} can be misleading and in part (or wholly) the
result of the net averaging of clouds with varying conditions. Even higher
signal-to-noise, and higher spectral resolving power, measurements of weak
absorption may be much more revealing about the carriers of the DIBs than the
strong DIBs found in the spectra of distant stars.

\begin{figure*}[t!]
\resizebox{\hsize}{!}{\includegraphics[clip=true]{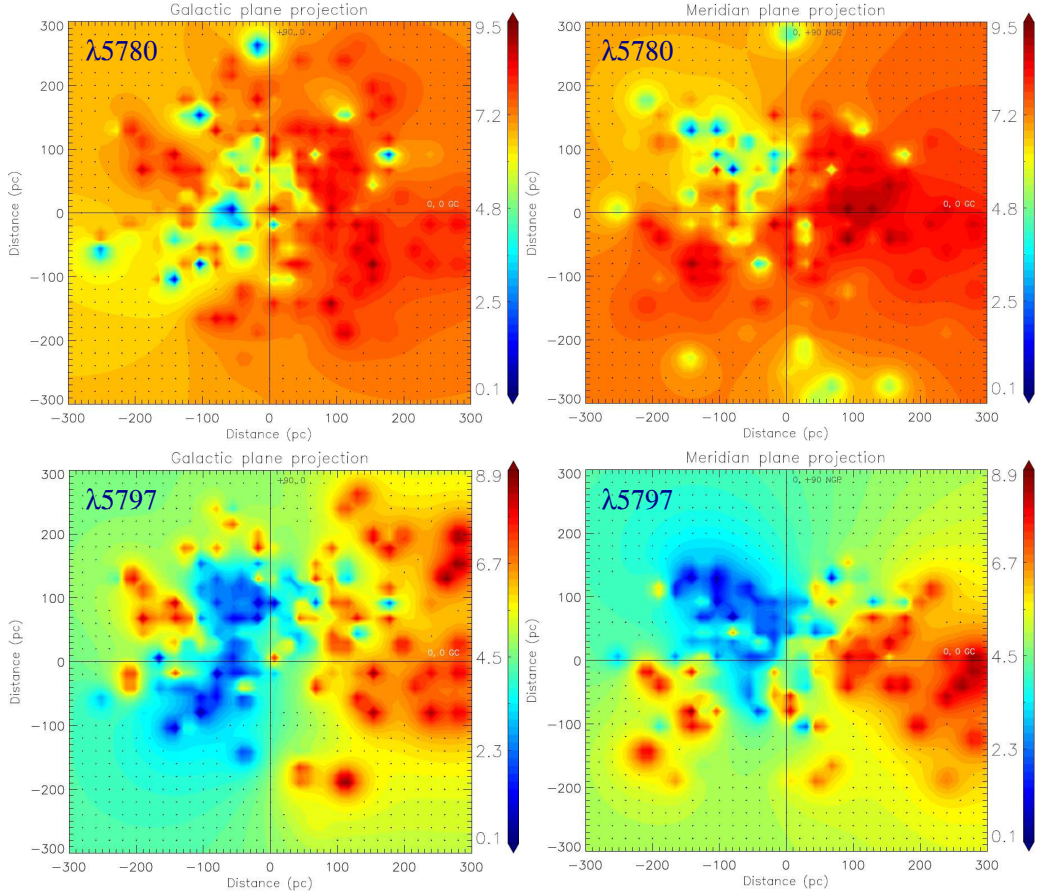}}
\caption{\footnotesize
Projections of the three-dimensional density (in arbitrary units) onto the
({\it left:}) Galactic plane and ({\it right:}) meridional plane perpendicular
to the former, of the ({\it top:}) $\lambda$5780 \AA\ and ({\it bottom:})
$\lambda$5797 \AA\ DIB. The DIB carriers are clearly depleted within a region
near to the Sun (located at the centre of these panels), but -- like in the
corresponding atomic maps -- there is evidence for small-scale structure and
for variations in the relative abundances of the (unknown) carriers of these
diffuse interstellar bands. Farhang et al.\ (2016).}
\label{3D}
\end{figure*}

The ultimate result of the survey is a three-dimensional map of the relative
density of the DIB carriers (figure~\ref{3D}; absolute values are not known as
the oscillator strength of the DIB transition remains unknown without the
carrier having been identified), using an inversion method
\citep{Vergely2001}. The method was tested on Na\,{\sc i} measurements from
the same survey data, and the resulting map compares well with the result by
\citet{Welsh2010} for a much larger sample of stars. In figure~\ref{3D} we
show two projections, for the $\lambda$5780 \AA\ and $\lambda$5797 \AA\ DIBs:
onto the Galactic mid-plane and onto the meridional plane perpendicular to the
Galactic plane. The former show the Local Bubble as traced by the DIBs
extending mostly away from the Galactic Centre direction (even though on this
scale, the Galactic Centre is some 25 times more distant than the extent of
the map, and it is unlikely that we are detecting a Galactic radial gradient
in density -- it is much more likely that there is a local spiral arm or spur
at that side of the Sun). The latter, the meridional projection shows the
Local Bubble opening into the extra-planar regions and possibly all the way
out into the Halo.

\begin{figure*}[t!]
\resizebox{\hsize}{!}{\includegraphics[clip=true]{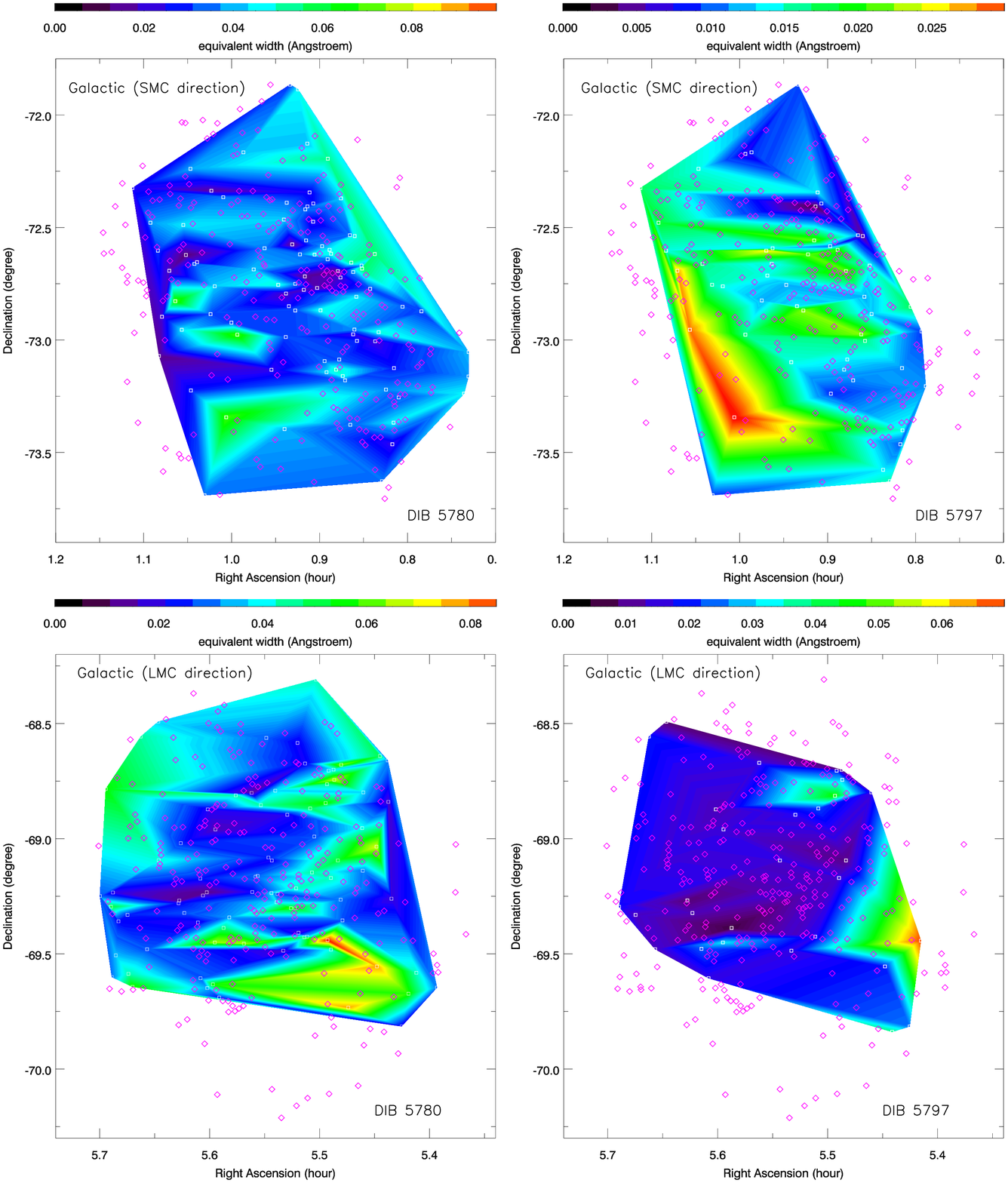}}
\caption{\footnotesize
Sub-degree variations across the sky in the Galactic foreground absorption in
front of the Small ({\it top}) and Large ({\it bottom}) Magellanic Clouds;
$\lambda$5780 \AA\ ({\it left}) and $\lambda$5797 \AA\ DIB ({\it right}).
\citet{Bailey2015a}.}
\label{Mag}
\end{figure*}

These three-dimensional maps already indicate a highly complex structure also
in the DIBs. Indeed, even smaller structures were found in degree-size maps,
towards extra-galactic targets \citep{vanLoon2009,vanLoon2013}. In
figure~\ref{Mag} we show for the first time maps of the Milky Way absorption
in the $\lambda$5780 \AA\ and $\lambda$5797 \AA\ DIBs as detected in the
foreground of over 300 stars in each of the Small and Large Magellanic Clouds
\citep{Bailey2015a}. Again, variations are seen on scales corresponding to
about a parsec if these structures are around 100 pc distant from us. And
again, these two DIBs do not seem to coincide very much.

\section{What's next?}

The surveys described here have demonstrated that it is possible to create
maps of DIBs across vast areas of galaxies also at low column densities where
we are most likely to see clear and true relations between -- and mutual
exclusions of -- different DIBs that can help us to identify their carriers.
The surveys have also highlighted the need for dedicated observations at much
higher signal-to-noise levels than what is typical in stellar or extragalactic
spectroscopic surveys (even if the latter often provide an opportunity for
interesting secondary science using DIBs). To distinguish different kinematic
components in the narrower DIBs, and help combat the interfering effects from
telluric and stellar photospheric spectral lines, high resolution (at least
$\lambda/\Delta\lambda>10\,000$) is highly desirable. Atomic and molecular
lines will become increasingly useful in such surveys. While the relatively
strong $\lambda$4428 \AA, $\lambda$5780 \AA, $\lambda$5797 \AA\ and
$\lambda$6614 \AA\ are the most obvious DIBs to target there are many other
DIBs both at optical and infrared wavelengths that could reveal information
left hidden by measurements of the above four DIBs, e.g.\ a triplet of DIBs in
the 470nm range \citep{vanLoon2013}.

We thus advocate the concept of a next generation of high resolution, high
signal-to-noise, broad range, multi-object spectroscopic surveys of thousands
of nearby stars and stars in the Magellanic Clouds. The requirements for these
surveys transcend current large spectroscopic programmes such as SDSS or
GAIA-ESO and even DIBs specific surveys such as EDIBLES at the ESO-VLT (PI:
Nick Cox). We note that the envisaged DIBs surveys would provide unprecedented
serendipitous scientific progress, in particular of an astrophysical nature.
They would need large investments of time at 4--10m class telescopes, but this
has become possible. They will also rely on accurate stellar atmosphere
modelling and spectral synthesis, but this too has become possible
\citep{Puspitarini2015}. The paybacks will be huge given that long-standing
and wide-ranging problems can finally be addressed accurately and
comprehensively. Apart from better understanding the multi-phase ISM, we will
know what the Solar System is moving through and we may finally find out what
material it is that causes DIBs.

\begin{acknowledgements}
We would like to thank Rosine Lallement and Ana Monreal-Ibero for organising
such an interesting and rather pleasant meeting. JvL acknowledges the award of
a Royal Society International Exchange grant (IE130487) and the award of a
travel bursary from Santander Bank, and he has realised that the Earth is
still spinning. MB acknowledges an STFC-funded studentship at Keele
University, where she had carried out the Southern hemisphere survey.
\end{acknowledgements}
\bibliographystyle{aa}

\begin{thebibliography}{}
\bibitem[{Bailey et al.\ (2015b)}]{Bailey2015b} Bailey, M., van Loon, J.~Th.,
Farhang, A., Javadi, A., Khosroshahi, H.~G., Sarre, P.~J., \& Smith, K.~T.\
2015b, \aap
\bibitem[{Bailey et al.\ (2015a)}]{Bailey2015a} Bailey, M., van Loon, J.~Th.,
Sarre, P.~J., \& Beckman, J.~E.\ 2015a, \mnras, 454, 4013
\bibitem[{Campbell et al.\ (2015)}]{Campbell2015} Campbell, E.~K., Holz, M.,
Gerlich, D., \& Maier, J.~P.\ 2015, \nat, 523, 322
\bibitem[{Farhang et al.\ (2015a)}]{Farhang2015a} Farhang, A., Khosroshahi,
H.~G., Javadi, A., et al.\ 2015a, \apj, 800, 64
\bibitem[{Farhang et al.\ (2015b)}]{Farhang2015b} Farhang, A., Khosroshahi,
H.~G., Javadi, A., \& van Loon, J.~Th.\ 2015b, \apjs, 216, 33
\bibitem[{Fitzgerald (1968)}]{Fitzgerald1968} Fitzgerald, M.~P.\ 1968, \aj,
73, 983
\bibitem[{Foing \& Ehrenfreund (1994)}]{Foing1994} Foing, B.~H., \&
Ehrenfreund, P.\ 1994, \nat, 369, 296
\bibitem[{Friedman et al.\ (2011)}]{Friedman2011} Friedman, S.~D., York,
D.~D., McCall, B.~J., et al.\ 2011, \apj, 727, 33
\bibitem[{Garc\'{\i}a-Hern\'andez \& D\'{\i}az-Luis (2013)}]
{GarciaHernandez2013} Garc\'{\i}a-Hern\'andez, D.~A., \& D\'{\i}az-Luis,
J.~J.\ 2013, \aap, 550, L6
\bibitem[{Heger (1922)}]{Heger1922} Heger, M.~L.\ 1922, Lick Observatory
Bulletin, 10, 141
\bibitem[{Herbig (1995)}]{Herbig1995} Herbig, G.~H.\ 1995, \araa, 33, 19
\bibitem[{Kos et al.\ (2014)}]{Kos2014} Kos, J., Zwitter, T., Wyse, R., et
al.\ 2014, Science, 345, 791
\bibitem[{Kre{\l}owski \& Westerlund (1988)}]{Krelowski1988} Kre{\l}owski, J.,
\& Westerlund, B.~E.\ 1988, \aap, 190, 339
\bibitem[{Lallement et al.\ (2003)}]{Lallement2003} Lallement, R., Welsh,
B.~Y., Vergely, J.-L., Crifo, F., \& Sfeir, D.\ 2003, \aap, 411, 447
\bibitem[{Lallement et al.\ (2014)}]{Lallement2014} Lallement, R., Vergely,
J.-L., Valette, B., Puspitarini, L., Eyer, L., \& Casagrande, L.\ 2014, \aap,
561, 91
\bibitem[{Lan, M\'enard \& Zhu (2015)}]{Lan2015} Lan, T.-W., M\'enard, B., \&
Zhu, G.\ 2015, \mnras, 452, 3629
\bibitem[{McCall et al.\ (2010)}]{McCall2010} McCall, B.~J., Drosback, M.~M.,
Thorburn, J.~A., et al.\ 2010, \apj, 708, 1628
\bibitem[{Puspitarini et al.\ (2015)}]{Puspitarini2015} Puspitarini, L.,
Lallement, R., Babusiaux, C., et al.\ 2015, \aap, 573, 35
\bibitem[{Redfield \& Linsky (2000)}]{Redfield2000} Redfield, S., \& Linsky,
J.~L.\ 2000, \apj, 534, 825
\bibitem[{Sarre (2006)}]{Sarre2006} Sarre, P.~J.\ 2006, Journal of Molecular
Spectroscopy, 238, 1
\bibitem[{Snowden et al.\ (2015)}]{Snowden2015} Snowden, S·~L., Koutroumpa,
D., Kuntz, K.~D., Lallement, R., \& Puspitarini, L.\ 2015, \apj, 806, 120
\bibitem[{van Loon et al.\ (2009)}]{vanLoon2009} van Loon, J.~Th., Smith,
K.~T., McDonald, I., Sarre, P.~J., Fossey, S.~J., \& Sharp, R.~G.\ 2009,
\mnras, 399, 195
\bibitem[{van Loon et al.\ (2013)}]{vanLoon2013} van Loon, J.~Th., Bailey, M.,
Tatton, B.L., et al.\ 2013, \aap, 550, 108
\bibitem[{Vergely et al.\ (2001)}]{Vergely2001} Vergely, J.-L., Freire
Ferrero, R., Siebert, A., \& Valette, B.\ 2001, \aap, 366, 1016
\bibitem[{Welsh et al.\ (2010)}]{Welsh2010} Welsh, B.~Y., Lallement, R.,
Vergely, J.-L., \& Raimond, S.\ 2010, \aap, 510, 54
\bibitem[{Zasowski et al.\ (2015)}]{Zasowski2015} Zasowski, G., M\'enard, B.,
Bizyaev, D., et al.\ 2015, \apj, 798, 35
\end{thebibliography}

\end{document}